\documentclass[aps,prl,showpacs,nobibnotes,twocolumn,
nobalancelastpage,superscriptaddress]{revtex4}
\usepackage{graphicx}
\usepackage{amsmath}
\usepackage{latexsym}
\usepackage{dcolumn}
\usepackage{bm}
\input{epsf}
\newcommand{\be}{\begin{equation}}
\newcommand{\ee}{\end{equation}}
\newcommand{\ba}{\begin{eqnarray}}
\newcommand{\ea}{\end{eqnarray}}
\newcommand{\bi}{\begin{itemize}}
\newcommand{\ei}{\end{itemize}}
\newcommand{\ga}{\gtrsim}
\newcommand{\bfi}{\begin{figure}
\epsfxsize=9cm
\epsffile}
\newcommand{\efi}{\end{figure}}

\newcommand{\la}{\lesssim}

\begin{document}
\title{The behavior of  $f(R)$ gravity in the solar system, galaxies
  and clusters}   
\author{Pengjie Zhang}
\email{pjzhang@shao.ac.cn}
\affiliation{Shanghai Astronomical Observatory, Chinese Academy of
  Science, 80 Nandan Road, Shanghai, China, 200030}
\affiliation{Joint Institute for Galaxy and Cosmology (JOINGC) of
SHAO and USTC}

\begin{abstract}
For cosmologically interesting $f(R)$ gravity models, we derive the
complete set of the linearized field equations in the 
Newtonian gauge, under environments of the solar system, galaxies and clusters
respectively.  Based on these equations, we confirmed previous
$\gamma=1/2$ solution in the solar system. However, $f(R)$
gravity models can be strongly environment-dependent and 
the high density  (comparing to the 
cosmological mean) solar system environment can excite a viable
$\gamma=1$ solution for some $f(R)$ gravity models. Although for
$f(R)\propto -1/R$, it is not the case; for
$f(R)\propto -\exp(-R/\lambda_2H_0^2)$, such $\gamma=1$
solution does exist. This
solution is virtually  indistinguishable 
from that in  general relativity (GR) and the value of the associated
curvature approaches 
the GR limit, which is much higher than value in the $\gamma=1/2$
solution. We show that for some forms of   
$f(R)$ gravity, this solution is physically stable in the solar system
and can smoothly connect to the surface of the Sun. 

The derived field equations can be applied directly to gravitational
lensing  of  galaxies and clusters. We find that, despite significant
difference in the environments of galaxies and
clusters comparing to that of the solar system, gravitational lensing
of galaxies and clusters  can be virtually  identical to that in GR,
for some forms of $f(R)$ gravity. Fortunately,  galaxy rotation
curve and  intra-cluster gas pressure profile  may contain valuable
information to distinguish these $f(R)$ gravity models from GR. 
\end{abstract}
\maketitle

\section{Introduction}
The standard theory of gravity (the general
relativity (GR)) combined with the standard model of particle physics
failed to explain a wide range of independent observations, from the expansion of the Universe, the cosmic microwave
background, the large  scale structure of the universe to galaxy and
cluster dynamics.  To reconcile
observations,  dark matter and dark energy, as modifications to particle
physics,  were proposed and work surprisingly well \cite{Reviews}. However, equally reasonable in logic, one can
modify  gravity instead to reconcile
observations.  It has been shown  that the modified Newtonian dynamics (MOND)
and its relativistic version Tensor-Vector-Scalar theory
\cite{MOND} can replace dark matter at galaxy scales to
reproduce galaxy rotation curves, and 5-D DGP 
gravity \cite{DGP} and $f(R)$ gravity \cite{fR1,fR} can replace dark
energy to reproduce the accelerated expansion of the universe. 

Like dark matter and dark energy, viable modifications in gravity must pass
all sorts of tests from the large scale structure of the universe
\cite{MONDLSS,DGPLSS,fRLSS,Song06,Bean06,Zhang06} to  galaxy and cluster dynamics to the
solar system  tests (SST). 
Unlike GR, which involves metric derivatives no higher than second
order, $f(R)$ gravity involves also third and fourth order
derivatives, which caused complications in the calculation
\cite{discussion}. 

 An outstanding question is whether $f(R)$ gravity is
consistent with SST, which have  put stringent
constraints on  the PPN parameter $\gamma=1\pm O(10^{-4})$
\cite{SST}.  In this parameterization, the 
Schwarzschild metric takes the form of $ds^2=(1-2GM/r+\cdots)dt^2-(1+2\gamma
GM/r+\cdots)dr^2+r^2d\Omega$. For GR, $\gamma=1$.
Various authors have
discussed the conditions for $f(R)$ gravity or  its extensions to pass
SST \cite{general}. Utilizing the equivalence between $f(R)$ gravity and a
special class of scalar-tensor theory, Chiba concluded that the
solution of 
$f(R)=-\mu^4/R$ gravity has  $\gamma=1/2$ and is thus ruled out by
SST \cite{Chiba03}. The equivalence between $f(R)$ gravity
and special  scalar-tensor theory  evoked some controversies
(e.g. \cite{Faraoni06,Faraoni07}). Controversies
also exist in approaches without resort to scalar-tensor theory. There
are different approaches to vary the $f(R)$ action to 
obtain the field equations. One is the metric formalism, in which the
only independent variable is the metric. Another is the Palatini form,
in which the connection is also an independent
variable. Based on the Palatini form, \cite{favor_Palatini}
concluded that $f(R)$ gravity  can in general be
perfectly consistent with SST. Same conclusion is reached in the
metric formalism by several groups
(e.g. \cite{Rajaraman03,favor3,Multamaki06}). However, based also on
the metric formalism, 
\cite{Erickcek06} confirmed the  $\gamma=1/2$ solution  for the
$f(R)=-\mu^4/R$ gravity. Furthermore, they pointed out that the
$\gamma=1$ vacuum solution with constant curvature can not connect
smoothly to the  surface of the Sun. This work has been extended to  a
wide range of  $f(R)$ gravity  models \cite{Chiba06}.  

To clarify this crucial issue, we derive the complete set of  linearized
field equations of the  two Newtonian 
potentials  $\phi$ and $\psi$, for cosmologically
interesting $f(R)$ gravity models, under the metric formalism. We draw
the attention that results in the Palatini form are in general   
different. The field equations turn out to take simple 
forms under the environments of the solar system, galaxies or
clusters. 
Based on these equations, we confirmed the $\gamma=1/2$ solution. However, we
also find that these equations accept the $\gamma=1$ solution for some
forms of $f(R)$ gravity, due to
the presence of non-negligible interplanetary dust and planets. 
This  large matter density in the solar system, when compared to the
cosmological mean, can significantly suppress the corrections induced to
GR by $f(R)$ gravity and reduces the field equations to the GR
limit. We further checked on the stability of the $\gamma=1$
solution and discussed ways to make it stable.

Galaxies and clusters are excellent laboratories to test $f(R)$
 gravity through gravitational lensing, cluster X-ray and SZ flux and
 galaxy rotation curve. We perturb the FRW background and derive the
 equations applicable  to galaxies and clusters. The derived equations
 can be applied directly to address the above observational
 properties. For some $f(R)$ gravity models, the  gravitational
 lensing effect is virtually identical to that  in  GR. However,
 galaxy rotation curve, cluster pressure 
 profile, the relation between the cluster mass, X-ray 
temperature, X-ray luminosity and the SZ flux, are modified, even for
 these $f(R)$  gravity models. These observations then have
 discriminating power for a wide range of $f(R)$ gravity models.

\section{Linearized field equations of $f(R)$ gravity}
The $f(R)$ gravity takes the action
\be
L=\int (R+f(R))\sqrt{-g}d^4x\ ,
\ee
and the field equation
\be
\label{eqn:field}
R_{uv}-\frac{1}{2}g_{uv}(R+f)+f_RR_{uv}+g_{uv}\Box f_R-f_{R;u;v}=8\pi
GT_{uv}\ .
\ee
Throughout the paper, we assume that $T_{\mu\nu}$ takes the form of ideal
fluid with negligible pressure. 
For  $f(R)$ gravity models of cosmological interest, the $f(R)$ term
must vanish at high redshifts and $R$ should approach its GR limit, in
order not to  conflict with early time  physics such as BBN and CMB.
At these high redshift the cosmological density is comparable to the solar
system local density where SST were carried out. So we would expect
that the correction induced by $f(R)$ can be significantly
suppressed.  However, this is only true if the GR limit, namely
$R\rightarrow 8\pi G\rho$, is reached. In this paper, we will
explicitly investigate on the feasibility of such solution. 

In the solar system, galaxies and  clusters, we expect that the  gravitational
field is  weak and the time variation of the field is negligible. Due to
different environments 
and different boundary conditions, the linearized field equations  in the
solar system differ slightly from that in galaxies and 
clusters. Thus we will treat the two cases separately. 

\subsection{Field equations applicable to the solar system}
In the solar system, we choose a static metric with the proper time 
\ba
ds^2&=&-g_{\mu\nu}dx^{\mu}dx^{\nu}\nonumber \\
&=&(1+2\psi({\bf
  x}))dt^2-(1+2\phi({\bf x}))\sum_i dx^{i,2}\ .
\ea
This is just the widely adopted Newtonian gauge in cosmology when
dropping the time dependence, where $\phi$ and $\psi$ are two Newtonian
potentials. Since $|\phi|,|\psi|\ll 1$, non-vanishing Ricci tensor components
are $R_{00}\simeq \nabla^2\psi$ and $
R_{ij}\simeq-\nabla^2\phi \delta_{ij}-(\phi+\psi)_{,ij}$. The curvature scalar
$R \simeq -2\nabla^2\psi-4\nabla^2\phi$.  In Eq. \ref{eqn:field},  it is safe
to  neglect  terms $(R+f)\phi$,  $(R+f)\psi$ with respect to
$R+f$ and neglect  terms $\phi\Box f_R$, $\psi \Box f_R$
with respect to  $\Box f_R$, since $\phi$, $\psi$ are small. Also, one can
approximate the covariant derivative 
 $f_{R;i;j}$ as the ordinary derivative $f_{R,ij}$, since
 $|\Gamma^{\sigma}_{ij}f_{R;\sigma}/f_{R,ij}|\sim |\phi|\ll 1$. 
We then obtain
\be
\label{eqn:00}
R_{00}(1+f_R)+\frac{1}{2}(R+f)-\Box f_R=8\pi G\rho\ ,
\ee
\be
\label{eqn:ii}
R_{ii}(1+f_R)-\frac{1}{2}(R+f)-\Box f_R-(f_R)_{,ii}=0\ ,
\ee
\be
\label{eqn:ij}
R_{ij}(1+f_R)-(f_R)_{,ij}=0\ \ {\rm when}\ i\neq j\ \ .
\ee

Before proceeding to the final results, we point out a generic constraint
exerted  by Eq. \ref{eqn:ij}, for constant curvature solutions. For this kind
of solutions, 
$f_{R;i;j}=f_{R,ij}=0$  and  
$(\phi+\psi)_{,ij}=0$.  Thus the coefficient of the $r^{-1}$ term in $\phi$ and
$\psi$  must be equal (with opposite sign). In another word,  {\it the
  constant curvature solution must have $\gamma=1$.} 

Since $|\phi+\psi|\ll 1$, Eq. \ref{eqn:ij} can be integrated to give 
\be
\label{eqn:phi_psi}
(\phi+\psi)(1+f_R)=-f_R+C_0r^2+{\rm const.}\ .
\ee
This result is straightforward to check. From 
Eq. \ref{eqn:phi_psi}, we get
$
(\phi+\psi)_{,ij}(1+f_R)+(\phi+\psi)f_{R,ij}+(\phi+\psi)_{,i}f_{R,j}+(\phi+\psi)_{,j}f_{R,i}=-f_{R,ij} 
\ .$
Since $|\phi+\psi|\ll 1$, the last three terms in the left hand side is
negligible comparing to the right hand side, we then obtain $
(\phi+\psi)_{,ij}(1+f_R)\simeq -f_{R,ij}$, 
This is the Eq. \ref{eqn:ij} to begin with. The integral would produce a term  $ax_i+bx_j$ in the right hand side of
Eq. \ref{eqn:phi_psi}. However  there is  no special
direction in the Universe, so it vanishes. The term $C_0r^2$ is
necessary. It reflects the
fact that, the flat Minkowski space-time is no longer the true background in
$f(R)$ gravity.

Combining Eq. \ref{eqn:00}, \ref{eqn:ii} and \ref{eqn:phi_psi}, we
obtain 
\be
\label{eqn:Poisson}
\nabla^2(\phi-\psi)=-\frac{8\pi G\rho+2C_0}{1+f_R}\ .
\ee
Eq. \ref{eqn:phi_psi} and \ref{eqn:Poisson} completely determine the
gravitational field of $f(R)$ gravity,
up to a constant $C_0$. $C_0$ can be determined by either
Eq. \ref{eqn:00}, \ref{eqn:ii}, or the trace of Eq. \ref{eqn:field}:
$(f_R-1)R-2f+3\Box f_R=-8\pi G\rho$. For example, when $f=\Lambda$ is the
cosmological constant, one can show $C_0=f/8$.

These equations do not require the condition of spherical
symmetry and can be applied to various environments including star forming
regions and star or black hole accretion disk. However, to
clarify the issue whether  $f(R)$ gravity is consistent with SST,  we apply
them to an idealized case, where  a point  
source with mass M (the Sun) is embedded in a background with
density $\rho_{\rm back}$ ($\rho=M\delta_D({\bf r})+\rho_{\rm back}$)
to sufficiently large radii. 

 The $\gamma=1/2$ solution
found in the literature \cite{Erickcek06,Chiba06} corresponds to a small
perturbation  induced by the 
central star (Sun) on top of the homogeneous and isotropic vacuum
background with curvature scalar $R_0$. For this solution, the
perturbation in the curvature 
scalar, is $R_1\simeq-2GM/3f_{RR}(R_0)r$, where $R_0$ is the curvature
scalar of the vacuum background.  The overall curvature scalar $R\equiv
R_0+R_1\ll 8\pi G\rho_{\rm back}$. Following the notation of
\cite{Sawicki07},  we call it the {\it low curvature
  solution}.   This $\gamma=1/2$ solution is not only viable for the
vacuum ($\rho_{\rm back}=0$, \cite{Erickcek06,Chiba06}) but also hold
for the realistic  configuration of $\rho_{\rm back}\neq 0$ 
around the Sun, as explicitly shown by \cite{Kainulainen07}.  One can
also  check that the $\gamma=1/2$ 
solution is indeed a solution of Eq. \ref{eqn:phi_psi} and
\ref{eqn:Poisson}. 

A crucial step to obtain the $\gamma=1/2$ solution starts with the
trace equation $(f_R-1)R-2f+3\Box f_R=-8\pi G\rho$. It can be 
rewritten in the form of $\Box f_R=-8\pi 
  G\rho/3+[2f-(f_R-1)R]/3$. In the limit that $8\pi
  G\rho\gg |2f-(f_R-1)R|$, one has $\Box f_R\simeq \nabla^2
  f_R=d^2f_R/dr^2+(2/r)df_R/dr=-8\pi 
  G\rho/3$. This is the key equation to reach the $\gamma=1/2$
  solution. We call the condition $8\pi
  G\rho\gg |2f-(f_R-1)R|$ as the low curvature condition. As
  explicitly shown in 
  \cite{Erickcek06,Chiba06,Kainulainen07}, the low curvature condition
  is satisfied for the $\gamma=1/2$
  solution. 

However, the trace equation can have another branch of
  solution.  It can be rewritten in the form of $R=8\pi
  G\rho+(f_RR-2f+3\Box f_R)$. In the limit that $|f_RR-2f+3\Box
  f_R|\ll 8\pi G \rho$, $R\simeq 8\pi
  G\rho$. We call it the {\it high curvature}
  solution and call the condition $|f_RR-2f+3\Box
  f_R|\ll 8\pi G \rho$ as the high curvature condition.  We will show
  that, given the fact that the local density 
  $\rho_{\rm back}$  is much higher  
than $\rho_c$ ($\rho_{\rm back}/\rho_c\ga 10^6$-$10^8$,
\cite{Zhang06}),  the high curvature condition can be satisfied and a
  viable $\gamma=1$ 
  solution is excited, for some forms of $f(R)$ gravity. The low
  curvature condition differ significantly from the high curvature
  condition \footnote{As seen from later discussion, the high
  curvature condition is equivalent to $|\Box f_R|\ll 8\pi G\rho$,
  while the low curvature condition is equivalent to $|\Box f_R|\simeq
  8\pi G\rho$.}.  Hence
  the existence of the $\gamma=1/2$ solution  does not invalidate 
  the existence  of the $\gamma=1$ solution.
Contrast to
  the $\gamma=1$ vacuum   
solution, the new $\gamma=1$ solution can connect smoothly to the
  surface of the Sun and can 
pass all SST. In the solar environment, it virtually reduces to GR,
  with $R\simeq 8\pi G\rho$.  For this solution, $f\rightarrow 0$ and $f_R
  \ll 1$.

 The solution of
  Eq. \ref{eqn:Poisson} under the spherical symmetry is  
\ba
\label{eqn:integral}
\phi-\psi&=&\frac{1}{1+f_R(r=0)}\frac{2GM}{r}\\
&&-\int dr\left[r^{-2} \int\left( \frac{8\pi G\rho_{\rm
  back}+2C_0}{1+f_R} r^2dr\right)\right] \nonumber\ .
\ea
Since now $|f_R|\ll 1$ we obtain 
\ba
\phi-\psi\simeq \frac{2GM}{r}-\int dr\left[r^{-2} \int\left(
  [8\pi G\rho_{\rm 
  back}+2C_0] r^2dr\right)\right] \nonumber\ .
\ea
Combining with Eq. \ref{eqn:phi_psi}, we obtain 
\ba
\label{eqn:solution}
\phi&=&\  \ \frac{GM}{r}-\int \frac{dr}{r^2}\int
  4\pi G\rho_{\rm 
  back} r^2dr+\frac{C_0r^2}{3}+{\rm const.}\ ,\nonumber\\
\psi&=& -\frac{GM}{r}+\int \frac{dr}{r^2}\int
  4\pi G\rho_{\rm 
  back} r^2dr+\frac{2C_0r^2}{3}+{\rm const.}\ ,\nonumber\\
\ea
where the constant $C_0$ is given by the trace of
  Eq. \ref{eqn:field}. By variable
transform  $r\rightarrow r^{'}=r(1+\phi)$, one can express 
the solution (Eq. \ref{eqn:solution}) in the form of the
Schwarzschild metric and then verifies the $\gamma=1$ result.

{\it Is this solution consistent with all approximations we made?} To
be specific, we discuss 
two forms of $f(R)$ gravity discussed in 
the literature,  
$f(R)=-\mu^4/R$ \cite{fR1} and $f(R)=-\lambda_1 H_0^2
\exp(-R/\lambda_2H_0^2)$ \cite{Zhang06}.  For $f(R)=-\mu^4/R$ to drive the
late time acceleration, 
$\mu\sim H_0$ where $H_0$ is the present day Hubble constant.  For
$f(R)=-\lambda_1 
H_0^2\exp(-R/\lambda_2H_0^2)$ proposed in \cite{Zhang06},
$\lambda_1\sim 1$ and $\lambda_2\sim 
10^3$ can produce virtually degenerate expansion rate to that of
$\Lambda$CDM cosmology. 

We first check the high curvature condition $|f_RR-2f+3\Box
  f_R|\ll 8\pi G \rho$. Since now $R\simeq 8\pi G\rho$, $|f_R|\ll 1$,
  the condition reduces to $|\Box f_R|\ll R\simeq 8\pi G\rho$. 
The high curvature condition requires
  $|d^3f/dR^3 R/r^2|\ll 1$ and $|d^2f/dR^2/r^2|\ll 1$ in general. Here, we
  have adopted the approximations $R^{'}\sim R/r$ and $R^{''}\sim R/r^2$
  where $r\neq 0$.  

 For $f(R)=-\mu^4/R$, $|d^2f/dR^2/r^2|\sim |[d^3f/dR^3][R/r^2]|\sim
  \mu^4/R^3r^2\sim (\rho_c/\rho)^3/(H_0^2r^2/c^2)\sim 10^{12} ({\rm 
  AU}/r)^2\gg 1$. Clearly, {\it $f(R)=-\mu^4/R$ model  does no accept the
  $\gamma=1$ solution and contradicts with SST.  }

For $f(R)=-\lambda_1 
H_0^2\exp(-R/\lambda_2H_0^2)$ and 
$\lambda_2=10^3$,  $\exp(-R/\lambda_2H_0^2)\sim
10^{-400}$. So $|d^2f/dR^2/r^2|\sim
\exp(-R/\lambda_2H_0^2)/(H_0^2r^2/c^2)\sim 10^{-370} ({\rm
  AU}/r)^2 \ll 1$. One can further check $|[d^3f/dR^3][R/r^2]|\ll 1$.
    {\it So for 
$f(R)=-\lambda_1   H_0^2\exp(-R/\lambda_2H_0^2)$, the high curvature
      condition is satisfied and $\gamma=1$ high curvature solution is
      accepted.}  Furthermore, the
weak field condition is also satisfied since 
$|\phi|,|\psi|\sim 10^{-8} [{\rm AU}/r]\ll 1$. The condition $|f_R|\ll 1$ is
perfectly satisfied too.  Hence-after, we only discuss this form of
$f(R)$ gravity, unless explicitly specified.

{\it Is it consistent with SST? Yes.} Since $C_0\ll
G\rho$, Eq. \ref{eqn:solution}  is
virtually identical to that in GR and causes no conflict with SST.
 Furthermore, in the solar system, 
$|\phi|,|\psi|\sim 10^{-8} (M/M_{\rm Sun})({\rm AU}/r)\gg |f_R|$ is generally
satisfied. $|C_0r^2/\phi|\ll 1$ is also satisfied. We then have
$\phi+\psi\simeq 0$ and $\nabla^2(\phi-\psi)\simeq -8\pi G\rho$. Thus the
field equations are virtually identical to that in GR and we should sense no
difference between $f(R)$ gravity and GR. 

This solution relies on the condition that $\rho_{\rm back}\gg
\rho_c$. Clearly, at sufficiently large distance, this condition is
violated. However, from Eq. \ref{eqn:integral} and the definition of
$\gamma$, it is clear that the 
value of $\gamma$ is determined by the behavior of $f_R$ at
$r\rightarrow 0$ instead of $r\rightarrow \infty$. So the $\gamma=1$
conclusion is not affected.

{\it Is it stable against small perturbations?} The stability of
solutions in  $f(R)$ gravity models has been widely discussed 
\cite{Dolgov03,Soussa04,Faraoni05,Song06,Bean06,Sawicki07}. For example,
\cite{Sawicki07} found that,  the high curvature
cosmological solutions may not be stable while the low curvature
solutions can in general be stable.  Here we briefly discuss the
stability of the two solar system solutions and leave the detailed
investigation elsewhere.

Perturbing the trace 
equation $(f_R-1)R-2f+3\Box f_R=-8\pi G\rho$ around $R_0=8\pi G\rho$,
we obtain $\ddot{R}_1+R_1/3f_{RR}(R_0)\simeq 0$. A negative $f_{RR}$
causes exponential amplification of the initial perturbation with a time scale
of 
$t_0=|3f_{RR}(R_0)|^{1/2}$.   For  $f(R)=-\lambda_1
H_0^2 \exp(-R/\lambda_2H_0^2)$, $f_{RR}(R_0)<0$ and the associated
$t_0\ll  1/H_0$, since $R_0\gg \lambda_2 H_0^2$. Thus the high
curvature solution is  highly unstable and 
thus unphysical for these two specific models. The low curvature
solution is unstable, too. However, the time scale $t_0$ is now
comparable to the Hubble time, since $R_0\sim H_0^2$. So the problem
of instability is much 
less severe. 

 However, for $f(R)=-\lambda_1 
H_0^2 \exp(-R/\lambda_2H_0^2)$, one can add another
small correction to make it stable in the solar system. We adopt a
form of $R^2$, 
which  was 
originally proposed by \cite{Starobinsky87} to drive inflation. Thus the
total   $f(R)=-\lambda_1  
H_0^2 \exp(-R/\lambda_2H_0^2)+\alpha R^2/H_0^2$, where $\alpha$ is
constant. This new form of 
$f(R)$ must satisfy the following conditions.  (1)  We  want to
preserve the high 
curvature condition. This requires $\alpha \ll (H_0r/c)^2\sim 10^{-30}
({\rm AU}/r)^2$. This condition automatically preserve
the condition $f_R(R_0)\ll 1$, which requires $\alpha\ll
\rho_c/\rho\sim 10^{-6}$.
(2) We require $f_{RR}(R_0)\geq
0$ to make the high curvature solution stable. This requires $\alpha \geq
\lambda_1\exp(-R_0^2/\lambda_2H_0^2)/2\lambda_2^2$. (3) Furthermore,
we want to preserve the merit of $f(R)$ 
gravity to drive the late time acceleration. This is satisfied if
$\lambda_1 H_0^2\exp(-R_{\rm FRW}/\lambda_2H_0^2)\gg \alpha R_{\rm
  FRW}^2/H_0^2$. This requires $\alpha \ll \lambda_1$ since the late
time FRW background curvature scalar $R_{\rm FRW}\sim H_0^2$. (4) We
want to preserve the success of early universe physics such as BBN and
CMB under GR and thus we want $\alpha R^2_{\rm FRW}(z_{\rm
  BBN})/H_0^2\ll R_{\rm FRW}(z_{\rm BBN})$.   BBN happened at $T\sim
    $ MeV and thus $z\sim 10^{10}$, where $R_{\rm FRW}(z_{\rm
      BBN})\sim 10^{30} H_0^2$. Thus we require $\alpha \ll
      10^{-30}$.  Since $\exp(-R_0/\lambda_2H_0^2)\sim 10^{-400}$, all these
conditions can be  satisfied.   Thus as
long as $\lambda_1 \exp(-R_0/\lambda_2H_0^2)/2\lambda_2^2 \leq \alpha \ll
10^{-30}$,  this form of $f(R)$ gravity contains a high curvature
solution, which is consistent with early and late time cosmological
observations of BBN, CMB, $H-z$ relation at $z\la 2$ and  SST, while
being physically stable in the solar system \footnote{The  cosmological 
  solution of $f(R)=-\lambda_1  
H_0^2 \exp(-R/\lambda_2H_0^2)+\alpha R^2/H_0^2$ with $\alpha\ll
10^{-30}$ is still unstable. Whether one can find suitable correction to $
f(R)=-\lambda_1   
H_0^2 \exp(-R/\lambda_2H_0^2)$ to make both the high curvature solar
system solution and the cosmological solution stable
requires further investigation. }.

\subsection{Field equations applicable to galaxies and clusters}
Galaxies and clusters are virialized objects embedded in the FRW
background. Approximately they are static in the physical
coordinate. So the time scale of the field variation is the Hubble
time and it is safe to neglect
all the time derivatives of $\phi$ and $\psi$. For a galaxy or
cluster at $z=1/a-1$, we choose a metric 
\ba
ds^2=(1+2\psi)dt^2-a^2(1+2\phi)\sum_i dx^{i,2}\ \ .
\ea
Eq. \ref{eqn:phi_psi} is then replaced by \cite{Zhang06}
\be
\label{eqn:phi_psi_cluster}
\phi+\psi=-f_R(R_{\rm FRW}+\delta R)+f_R(R_{\rm FRW})\ .
\ee
Here, $R=R_{\rm FRW}+\delta
R=6(\dot{a}^2/a^2+\ddot{a}/a)-2\nabla^2\psi-4\nabla^2\phi$. Throughout this
section, the
derivative is with respect to the physical coordinate.  The term $C_0 r^2$ presented in the solar system
solution  vanishes, because now the FRW background is the right background.
One can 
simply verify $C_0=0$ by the boundary condition that when $r\rightarrow
\infty$, $\phi\rightarrow 0$, $\psi\rightarrow 0$ and $R\rightarrow R_{\rm
  FRW}$.

The Poisson equation (Eq. \ref{eqn:Poisson}) is replaced by
\be
\label{eqn:Poisson_cluster}
\nabla^2(\phi-\psi)=-\frac{8\pi G(\rho_m-\bar{\rho}_b)}{1+f_R}\ .
\ee
Here, $\rho_m$ is the matter density of galaxies or clusters and
$\bar{\rho}_b$ is the background matter density.   
These equations  have been derived under the
quasi-static approximation at sub-horizon scales \cite{Zhang06}.
Equations applicable to galaxies and clusters are very
similar. A good thing is that now the quasi-static approximation is well
satisfied at all relevant scales in galaxy and cluster environments. 

The gravitational lensing is governed by the combination
$\phi-\psi$. So, given the same matter distribution, the gravitational
lensing effect of a  galaxy or a cluster in $f(R)$ gravity is
identical to that in GR, except a change in the Newton's constant from
$G$ to $G/(1+f_R)$. For $f(R)$ to drive late time acceleration,
$f_R>0$ in general, thus the lensing signal will be smaller by a
factor $f_R/(1+f_R)$. However, for some $f(R)$ gravity
models, such as $f(R)=-\lambda_1  
H_0^2 \exp(-R/\lambda_2H_0^2)+\alpha R^2/H_0^2$ proposed above, 
$|f_R|\ll 1$ in galaxy  
and cluster environments, so the difference may not be observable. 

However, galaxy rotation curve and intra-cluster gas pressure
profile can be significantly different to that in GR. The acceleration
of a test particle is $\dot{\bf v}=-\nabla  \psi$. The gas pressure $p$ is
determined by $\nabla p=-\rho \nabla \psi$. The matter density decreases from
$\sim 10^4 \rho_c$ close to the center to $\sim 40 \rho_c$ at virial
radius. The 
corresponding variation in $f_R$ is then comparable to variations in
$\phi,\psi$ and thus $\phi+\psi=0$ no 
longer holds.  So in galaxy
and cluster environments, $\psi$ in $f(R)$ gravity does not follow the Poisson
equation, as that in GR does.   Whether the difference caused is observable
is currently under investigation.

\section{Summary}
To investigate  $f(R)$ gravity in the solar system, galaxies and clusters, 
we derive the complete sets of the field equations which determine the two
Newtonian potentials $\phi$ and $\psi$, under corresponding
environments. In the  solar system, we found that some  $f(R)$
gravity models of cosmological   interest do contain physically stable
solutions which are virtually indistinguishable from GR. 

We predict that gravitational
lensing effect of quasi-static celestial objects such as 
galaxy and clusters in  $f(R)$ gravity is virtually the same  as in
GR, up to a change in the Newton's constant. However, 
galaxy rotation curve and  cluster gas pressure profile differ
intrinsically from that in GR. Thus observations of galaxies and cluster
dynamics are promising to put useful constraints on $f(R)$ gravity. 

{\bf Acknowledgments.}---
We  thank Adrienne Erickcek and  Tristan Smith for useful
discussions. PJZ is supported  by the One-Hundred-Talent Program of
the   Chinese Academy of Science, the NSFC grant 10533030 and
the CAS grant KJCX3-SYW-N2.

\end{document}